\documentclass[sigconf]{acmart}

\usepackage[super]{nth}
\usepackage{amsfonts}
\usepackage{dsfont}
\usepackage{enumitem}
\usepackage{makecell}
\usepackage{multirow}
\usepackage{booktabs}
\usepackage{graphicx}
\usepackage{subcaption}
\usepackage{xspace}
\usepackage{flushend}
\usepackage{arydshln}
\usepackage{pifont}
\usepackage{balance}
\usepackage[skip=2pt]{caption} 

\newcommand{\cmark}{\ding{51}}
\newcommand{\xmark}{\ding{55}}

\usepackage{array}
\usepackage{tikz}
\usetikzlibrary{tikzmark, arrows.meta, positioning}

\definecolor{proColor}{RGB}{34,139,34}
\definecolor{antiColor}{RGB}{178,34,34}
\definecolor{lineColor}{RGB}{140,140,140}

\copyrightyear{2026}
\acmYear{2026}
\setcopyright{acmlicensed}

\acmConference[WebSci '26]{18th ACM Web Science Conference}{May 26--29, 2026}{Braunschweig, Germany}
\acmBooktitle{18th ACM Web Science Conference (WebSci '26), May 26--29, 2026, Braunschweig, Germany}
\acmDOI{10.1145/3795766.3799768}
\acmISBN{979-8-4007-2504-3/2026/05}

\begin{document}

\title[Who Shapes Brazil's Vaccine Debate?]{Who Shapes Brazil’s Vaccine Debate? Semi-Supervised Modeling of Stance and Polarization in YouTube’s Media Ecosystem}

\author{Geovana S. de Oliveira}
\affiliation{
  \institution{Universidade Federal de Ouro Preto}
  \city{João Monlevade}
  \country{Brazil}
}
\email{geovana.so@aluno.ufop.edu.br}

\author{Ana P. C. Silva}
\affiliation{
  \institution{Universidade Federal de Minas Gerais}
  \city{Belo Horizonte}
  \country{Brazil}
}
\email{ana.coutosilva@dcc.ufmg.br}

\author{Fabricio Murai}
\affiliation{
  \institution{Worcester Polytechnic Institute}
  \city{Worcester}
  \country{United States}
}
\email{fmurai@wpi.edu}

\author{Carlos H. G. Ferreira}
\affiliation{
  \institution{Universidade Federal de Ouro Preto}
  \city{Ouro Preto}
  \country{Brazil}
}
\email{chgferreira@ufop.edu.br}

\renewcommand{\shortauthors}{Oliveira, et al.}

\begin{abstract}
Vaccination remains a cornerstone of global public health, yet the COVID-19 pandemic exposed how online misinformation, political polarization, and declining institutional trust can undermine immunization efforts. Most of the prior computational studies that analyzed vaccine discourse on social platforms focus on English-language data, specific vaccines, or short time windows, impairing our understanding of long-term dynamics in high-impact, non-English contexts like Brazil, home to one of the world’s most comprehensive immunization systems. We here present the largest longitudinal study of Brazil’s vaccine discourse on YouTube, leveraging a semi-supervised stance detection framework that combines self-labeling and self-training to classify nearly 1.4 million comments. By integrating stance with temporal patterns, engagement metrics, and channel taxonomy (legacy media, science communicators, digital-native outlets), we map how pro- and anti-vaccine narratives evolve and circulate within a hybrid media ecosystem. Our results show that semi-supervised learning substantially improves stance classification robustness, enabling fine-grained tracking of public attitudes across Brazil’s full immunization schedule. Polarization spikes during epidemiological crises, especially COVID-19, but becomes fragmented across vaccines and interaction patterns in the post-pandemic period. Notably, science communication and digital-native channels emerge as the primary loci of both supportive and oppositional engagement, revealing structural vulnerabilities in contemporary health communication. Thus, our work advances computational methods for large-scale stance modeling while offering actionable evidence for public health agencies, platform governance, and online information ecosystems.
\end{abstract}



\maketitle

\section{Introduction}

Vaccination remains one of the most effective public health interventions ever developed, credited with preventing millions of deaths annually and significantly reducing the burden of infectious diseases worldwide, as consistently documented by the World Health Organization \cite{WHO:2022, Fallucca:2024}. Historically, fluctuations in vaccine uptake have often followed large-scale epidemiological events, such as measles, yellow fever, and influenza outbreaks, which periodically realign public attention and reshape risk perception \cite{Pan:2017, Casey:2019, Barbier:2022}. The COVID-19 pandemic intensified these dynamics on an unprecedented scale, exposing societies to simultaneous challenges of scientific uncertainty, rapidly evolving policies, and massive surges of digital communication \cite{Mitts:2022, Zhao:2023}. As vaccination campaigns expanded worldwide, online platforms became central arenas for public deliberation, contestation, and the circulation of health information \cite{xue:2020, Sufi:2022}.

At the same time, the rise of vaccine hesitancy, politicization, and coordinated misinformation has emerged as a critical global challenge \cite{Alsabban:2021, malagoli:2021, Larson:2022}. Several studies have documented how fear-based narratives, conspiracy theories, and distrust in institutions proliferate online, influencing risk perception and undermining public health interventions \cite{Mitts:2022, aygun:2022, ferreira:2022}. Research across platforms shows that anti-vaccine discourse tends to cluster around emotionally charged topics such as adverse events, government mandates, and pharmaceutical skepticism, and often features distinct linguistic markers, like anger, anxiety, and moralized language \cite{malagoli:2021, straton:2023}. The pandemic exacerbated these patterns, intertwining vaccine uptake with political identity, misinformation ecosystems, and affective polarization \cite{melton:2022, hwang:2022}.

A growing computational literature has examined these phenomena on many platforms, including Twitter, Reddit, and YouTube. Studies analyze temporal variation in public sentiment \cite{qorib:2023, melton:2021, malagoli2021caracterizaccao, malagoli:2021}, explore aspect-specific perceptions such as political, health, and media frames \cite{aygun:2022}, and model behavioral constructs informed by psychological theory, including the Health Belief Model and the Theory of Planned Behavior \cite{jingcheng:2020}. Cross-platform comparisons reveal systematic differences in negativity, adversarial engagement, and toxicity \cite{melton:2022, huang:2024}. Research also highlights the role of platform algorithms in amplifying certain narratives and facilitating the spread of anti-vaccine content \cite{song:2017, lahouati:2020}. For YouTube specifically, studies find that anti-vaccine videos often achieve greater visibility, centrality, and engagement than pro-vaccine ones, even under moderation  \cite{locatelli:2022,abao:2021,miyazaki:2024,tokojima:2020}.

Across this body of work, however, two important limitations emerge. First, most computational studies focus on English-language content or on a narrow subset of vaccines, typically COVID-19, Human Papillomavirus (HPV), or Measles, Mumps, and Rubella (MMR), limiting our understanding of how public attention and stance evolve across a full national immunization schedule. Second, while previous research documents platform-level sentiment trends, few studies examine the structural organization of stance in YouTube’s information ecosystem, specifically, how different types of channels (news media, influencers, science communicators) shape the spread, visibility, and persistence of vaccine related narratives over time. This gap is particularly pronounced for Brazil, a country of continental scale whose the National Immunization Program (PNI) is widely recognized by the World Health Organization as one of the most comprehensive and successful immunization systems globally \cite{Silva:2020, Percio2023}. Yet, Brazil also experienced one of the most politically charged and polarized vaccine debates during the COVID-19 pandemic, marked by institutional contestation, distrust in public authorities, and large-scale circulation of misinformation \cite{locatelli:2022, tokojima:2020, malagoli:2021}. Understanding how Brazil’s vaccine discourse evolved across pre-pandemic, pandemic, and post-pandemic periods, especially within YouTube, the country’s most widely used platform for news consumption \cite{IBOPE_2023}, is essential for situating contemporary challenges in vaccination uptake and institutional trust.

In light of this, this paper aims to investigate how public discourse and polarization around vaccination evolved in Brazil’s YouTube ecosystem from 2018 to 2024, notably covering the COVID-19 pandemic, by developing and applying \textit{semi-supervised stance detection models} to characterize shifts in user attitudes, engagement dynamics, and content ecosystems across the full brazilian immunization schedule. We address this goal through three research questions:
\begin{description}[leftmargin=0pt,labelsep=0.5em,itemsep=4pt]
    \item[\textbf{RQ1}] \textit{How can semi-supervised approaches such as self-labeling and self-training improve stance detection robustness in large-scale YouTube discussions on vaccination?}

    \item[\textbf{RQ2}] \textit{How did the volume, stance distribution, and intensity of polarization in vaccination related discourse on YouTube shift across pre-pandemic, pandemic, and post-pandemic phases, and how do these shifts manifest across vaccines in the Brazilian PNI?}

    \item[\textbf{RQ3}] \textit{Which types of YouTube channels, such as news media, alternative media, influencers, and science communicators, most effectively amplify pro- and anti-vaccine narratives, and how do engagement dynamics shape the visibility and persistence of these narratives?}
\end{description}

To answer these questions, we compile a large longitudinal dataset of Brazil’s vaccine discourse on YouTube, covering seven years, multiple vaccines from the PNI, and millions of user comments. We then develop a semi-supervised pipeline based on self-labeling and self-training to classify stance across highly imbalanced data and sparse supervision signals. We integrate stance predictions with channel typologies, engagement metrics, and temporal features to characterize how polarization emerges and circulates across the platform. Our findings reveal three overarching insights. First, semi-supervised methods substantially improve stance detection accuracy in text classification settings characterized by ambiguity and extreme class imbalance. Second, vaccine discourse exhibits sharp temporal reconfiguration during the pandemic, with crisis-driven concentration followed by post-pandemic fragmentation. Third, polarization is anchored not primarily in legacy journalism, but in a hybrid media ecosystem dominated by science communicators and digital-native channels. For transparency and to support reproducibility, we release our stance model for community use\footnote{\url{https://huggingface.co/gseovana/llama-vaccine-stance-ptbr-lora}}.

\section{Related work}\label{sec:related-work}

Research on vaccination discourse across digital platforms spans three main domains: sentiment and hesitancy, polarization and information flows, and stance modeling. Across these areas, studies show that online vaccine debates are highly sensitive to contextual shocks, shaped by platform affordances, and unevenly distributed across media ecosystems.

\textit{Sentiment trends and hesitancy.} Early studies, often based on Twitter, tracked how public attitudes responded to rollouts, safety announcements, and outbreaks \cite{malagoli:2021,qorib:2023}. These studies identify strong correlations between external events and fluctuations in optimism, fear, and anger, with negative sentiment clustering around uncertainty, adverse effects, and politicized narratives. Aspect-based analyses further show that political framing frequently outweighs biomedical risk as a driver of negativity \cite{aygun:2022}. Cross-platform comparisons indicate that Reddit discussions are generally less negative and polarized than YouTube or Twitter, likely due to distinct moderation norms \cite{huang:2024,melton:2022}, though concerns about side effects and long-term risks remain consistent \cite{melton:2021}. Linguistic and emotional analyses highlight the role of anticipation, fear, anger, and anxiety in shaping engagement dynamics, often coupled with high-arousal misinformation and skepticism \cite{xue:2020,ferreira:2022,romy:2022}.

\textit{Polarization, toxicity, and information flows.}
A second strand examines how platform structure and recommendation systems shape polarization, particularly on YouTube. Negative or conspiratorial content often achieves higher centrality and algorithmic reinforcement \cite{song:2017}, while toxic comments can trigger contagious emotional escalation \cite{miyazaki:2024}. Comparative studies show that anti-vaccine ecosystems differ across national contexts, with Brazilian content exhibiting lower toxicity and engagement than U.S.\ counterparts \cite{locatelli:2022}. Additional work documents persistent disinformation linking vaccines to autism and the monetization structures that sustain such content \cite{tokojima:2020}. Network analyses further demonstrate that polarized clusters amplify distinct informational sources, with pro-vaccine users promoting scientific and journalistic outlets, and anti-vaccine groups circulating pseudoscientific and alternative-health domains \cite{monsted:2022}. Emotional responses also vary by content type, with supportive videos evoking joy and anticipation, and anti-vaccine content eliciting sadness and fear \cite{abao:2021}.

\textit{Automated stance classification}. Most prior stance studies rely on relatively small annotated datasets. Traditional classifiers show limited performance, particularly with sarcasm or implicit positions \cite{kunneman:2020}. Deep learning models, including multilingual BERT, CNN-based architectures, and hybrid approaches, substantially improve accuracy \cite{straton:2023,blanco:2025}. However, these approaches remain constrained by annotation costs, domain mismatch, and limited scalability to multi-million comment corpora or multi-year analyses.

Despite these advances, prior work typically focuses on narrow time windows, often centered on COVID-19, and rarely integrates multiple vaccines, platform-specific media structures, and stance dynamics within a unified longitudinal framework. These limitations are especially salient in Brazil, a continent-scale country with one of the world’s most comprehensive National Immunization Programs~\cite{Percio2023}. We address these gaps by combining a scalable semi-supervised stance pipeline with a seven-year longitudinal analysis covering diverse vaccines and linking stance patterns to channel typology, journalistic certification, and engagement dynamics. Together, our contributions provide a comprehensive mapping of how Brazil’s vaccine debate is structured, who shapes it, and how YouTube’s hybrid information ecosystem mediates public health communication over time.

\section{Methodology}\label{sec:methodology}
This section presents our methodology, combining data collection, semi-supervised stance detection, and sociotechnical analysis to study the Brazilian vaccination debate on YouTube over 7 years.

\subsection{Data Collection and Preprocessing}

To construct the dataset used in this study, we employ the YouTube Data API v3\footnote{\url{https://developers.google.com/youtube/v3/getting-started}}, which enables the automated retrieval of videos, channels, metadata and engagement statistics (comments, likes, and views). We set the API to return content in Portuguese, geolocated to Brazil, ensuring that the collected material reflects the Brazilian vaccination information ecosystem, following prior works \cite{costa2025characterizing, da2024monitorando, tanure2025caracterizaccao, dias2024analise}.

We begin the data collection by constructing a manually curated set of query terms focused on the Brazilian vaccination context. These terms derive from the Brazilian National Immunization Program (PNI) and official terminology from the Ministry of Health, as well as keywords associated with vaccine-preventable diseases. To ensure terminological consistency, we incorporate standardized vaccine and disease nomenclature, as well as immunization policy terms provided by DataSUS \cite{datasus}---the national public health information system responsible for maintaining epidemiological, immunization, and healthcare records across Brazil. In total, the PNI covers 30 vaccines. However, our analysis concentrates solely on the 19 vaccines included in the Brazilian National Immunization Schedule, prioritizing the discourse surrounding widely distributed and mandatory doses. The complete list of keywords and search queries is provided in Appendix \ref{appendix:search_strategy}.

We define the temporal scope of the dataset as January 2018 to July 2024. This period was chosen to capture the evolution of public debate before, during, and after the COVID-19 pandemic, including potential influences of national political events. For each query term, and for each month within this interval, we retrieve all metadata and comments from the top 50 most relevant videos returned by YouTube’s internal ranking system. During exploratory inspection, we identify substantial noise in the collected data. Word cloud analyses of video titles reveal a subset of educational and exam preparation content, which, although containing vaccine related terms, does not reflect public discussion or user stance. To remove such irrelevant material and preserve thematic coherence, we automatically exclude videos whose titles contain Portuguese educational terms like \textit{``professor'', ``aula'' (video lesson), ``curso'' (course), ``prova'' (exam),} and \textit{``revisão'' (content review)}.

Next, we restrict the dataset to comments written in Portuguese, as many videos attract multilingual engagement. To ensure high-precision language identification on short and informal texts, we combine a fine-tuned RoBERTa-base model\footnote{\url{https://huggingface.co/papluca/xlm-roberta-base-language-detection}}
 with the \texttt{LanguageDetection} library\footnote{\url{https://pypi.org/project/langdetect/}}, leveraging both context-aware predictions and robustness to short inputs \cite{Zola:2020}. A comment is retained as Portuguese if identified as such by at least one detector, reducing false negatives. After preprocessing, the final dataset comprises 3,897 channels, 14,318 videos, and 1,422,406 comments from 591,760 unique users, forming the basis for subsequent stance modeling and discourse analysis.

\subsection{Stance Classification Framework (RQ1)}

In this section, we describe the methodological framework used to build our stance classification system.

\subsubsection{Manual Annotation Protocol}

The manual annotation step is central to building a reference dataset for supervised training and model validation. Because annotating the full corpus is infeasible, we adopt a temporal sampling strategy in which we randomly select 50 comments from each month between 2018 and 2024. This approach prioritizes thematic diversity and temporal representativeness, capturing fluctuations in public discourse across vaccines, health events, and sociopolitical contexts.

Three independent annotators participated in the labeling process. All annotators were undergraduate students in Information Systems, aged between 21 and 25 at the time of annotation. To minimize individual biases, annotators worked independently and followed a predefined labeling protocol, where each stance was assigned at the comment level based solely on the textual content, without incorporating video-level information or broader conversational context. Each comment was assigned to one of three mutually exclusive categories: \textit{Favorable}, \textit{Against}, or \textit{Inconclusive}. The guidelines for each category are defined in the Appendix \ref{appendix:guidelines}.

We assess inter-rater agreement using Fleiss’ Kappa \cite{kappa:1971, fleiss:1981}, a widely used metric for multi-annotator categorical agreement that accounts for chance. 
The annotation process achieves $\kappa = 0.69$ (substantial agreement \cite{landis:1977}). Final labels are assigned by majority vote. In cases of complete disagreement, where each annotator selects a different category, the comment is deemed too ambiguous for reliable classification. Only 23 comments fall into this situation and are removed due to their negligible impact on the dataset. After completing this procedure, the validated labeled set contains 3,476 comments: 295 labeled as \textit{Favorable}, 446 as \textit{Against}, and 2,735 as \textit{Inconclusive}. This distribution reflects the natural imbalance commonly observed in real-world social media discussions about vaccination. Although not an artifact of sampling or annotation, this imbalance poses challenges for supervised learning and motivates the semi-supervised strategy described later.

\subsubsection{Supervised Model Training and Validation}\label{subsec:model}

The classification model we use in this work is Llama 3.1 8B \cite{dubey:2024}, an open source Large Language Model (LLM) based on the Transformer architecture. Its self-attention mechanism captures long-range dependencies and subtle discursive cues, which are important for informal online discourse containing irony, ambiguity, and colloquial language.

To adapt the model to the stance detection task, we perform supervised parameter-efficient fine-tuning using the manually annotated dataset. Specifically, we adopt Quantized Low-Rank Adaptation (QLoRA) to substantially reduce memory requirements \cite{dettmers:2023} by freezing the pretrained model weights and injecting low-rank adaptation matrices into selected components of the architecture, such as the attention projection layers (\texttt{q\_proj}, \texttt{k\_proj}, \texttt{v\_proj}). 

We quantize the model using the 4-bit NF4 scheme with \texttt{bfloat16} computation and adapt it for sequence classification through QLoRA modules (rank = 64, $\alpha$ = 16). The context length is set to 192 tokens, which covers more than 95\% of all comments without truncation. We add a shallow, task-specific classification head on top of the frozen Llama backbone, consisting of a single linear layer that maps the final hidden representation into a $C$-dimensional logits vector (one per class), followed by a softmax function. 
We set the learning rate to $2 \times 10^{-4}$ with no dropout, following previous work in the literature~\cite{Fonseca:2025}. We train the model with a batch size of 128 for up to 20 epochs using mixed-precision training (FP16). To address the strong class imbalance present in the dataset, especially the predominance of the \textit{Inconclusive} class, we employ a weighted cross-entropy loss in which class weights are set inversely proportional to class frequency and normalized to sum to 1. This design ensures that all classes contribute in the same proportion to the loss, preventing the model from converging to always predict the majority class (\textit{Inconclusive}). 

We use stratified 5-fold cross-validation, explicitly separating training, validation, and test sets. At each iteration, three folds are used to train the model, one fold is reserved for hyperparameter selection (validation), and a different fold is held out for final evaluation (test). Across iterations, all folds rotate through these roles. We use early stopping, with a patience of three epochs, to mitigate overfitting by halting training when validation performance no longer improves. We define the best checkpoint for each fold as the model that achieves the highest Macro F1 score on the validation set, a more informative metric for imbalanced multi-class scenarios~\cite{Bengio:2017}. Finally, we evaluate the model performance using Accuracy, Precision, Recall, F1-score, and Macro F1. 
All experiments were conducted on a server equipped with an Intel Xeon Gold 6442Y 2.6 GHz, a NVIDIA A40 GPU with 48 GB of VRAM, and 512 GB of RAM.

\subsubsection{Semi-supervised Learning via Self-Training}

To further improve stance classification performance, we incorporate a semi-supervised self-training strategy. This approach addresses two key challenges: severe class imbalance and high manual labeling costs. In self-training, the model initially trained on the labeled data (the baseline classifier) assigns pseudo-labels to the large pool of unlabeled comments. Selected pseudo-labeled examples are then incorporated back into the training set, allowing the model to benefit from additional supervision without requiring manual annotation \cite{Amini:2025}.

However, indiscriminately adding pseudo-labeled examples is a double-edged sword. Although increasing the training set often improves generalization, it can also amplify model errors, reinforce existing biases, and introduce semantic drift, particularly when pseudo-labels are uncertain or disproportionately belong to majority classes~\cite{Gui:2024}. To mitigate these risks, we adopt a confidence-based selection strategy grounded in predictive uncertainty. Specifically, we use Shannon entropy to quantify the uncertainty of each prediction on the unlabeled dataset~\cite{Gray:2011}. For a classification task with $C$ classes and predicted probability vector $\mathbf{p} = (p_1, p_2, \dots, p_C)$, entropy is defined as \( H(\mathbf{p}) = -\sum_{i=1}^C p_i \log p_i \). 
Entropy provides a smooth measure of how concentrated or diffuse the prediction is. Intuitively, low entropy corresponds to sharply peaked probability distributions in which the model strongly favors a single class, while high entropy reflects uncertainty or ambiguous evidence.

Given the scale of the unlabeled dataset (over 1.3M comments) and its substantial skew toward the \textit{Inconclusive} class, we adopt a low-entropy sampling strategy \cite{Amini:2025}, selecting only predictions for which the model demonstrates high confidence. First, the classifier fine-tuned on the initial labeled dataset 
generates probability distributions over the three stance classes for all unlabeled comments. Shannon entropy is then computed for each prediction probability distribution to quantify its level of uncertainty. Next, we select only comments falling below a percentile derived from the empirical entropy distribution. This selection strategy relies on minimizing entropy, which corresponds to retaining predictions with extremely high confidence. 

To counterbalance the natural dominance of the \textit{Inconclusive} class, this selection process incorporates inverse-frequency sampling, giving priority to high-confidence predictions from minority classes (\textit{Favorable} and \textit{Against}) by determining sample sizes proportional to the inverse class frequency ($1/N_c$) and selecting the corresponding lowest-entropy predictions. The resulting pseudo-labeled examples are then added to the training dataset. Finally, we train a new model on this expanded dataset. Since the extended distribution remained imbalanced, we recomputed the inverse-frequency weights based on the new class counts to calculate the weighted loss, ensuring the model focuses on minority classes during this second fine-tuning stage. The expanded dataset, composed of manually annotated samples and high-confidence pseudo-labeled comments, is then used to train a second-stage classifier. This semi-supervised approach improves overall performance and, critically, enhances generalization on underrepresented classes. The performance comparison before and after dataset enrichment is presented in the Results section.

\subsection{Large-Scale Stance Inference and Discourse Characterization (RQ2)}

After training and validating the stance classification model, we apply it to the full corpus of more than 1.3 million comments, obtaining for each comment a predicted stance label (\textit{Favorable}, \textit{Against}, or \textit{Inconclusive}). These labels form the basis for the large-scale analyses used to address RQ2, characterizing how vaccination discourse evolves over time and across vaccines in the Brazilian National Immunization Program.

To examine temporal dynamics, we segment the timeline into three analytically meaningful phases: pre-pandemic (Jan 1, 2018--Mar 10, 2020), pandemic (Mar 11, 2020--May 04, 2023), and post-pandemic (May 05, 2023--July 1, 2024) following official World Health Organization (WHO) declarations and the timeline of vaccination rollout in Brazil\footnote{\url{https://www.paho.org/en/news/2-7-2020-timeline-whos-response-covid-19}}. Within these exact boundaries, we derive summary statistics, such as the mean proportion of each stance category.

To analyze vaccine specific discourse, we link the stance-classified comments to the vaccines included in the Brazilian National Immunization Schedule\footnote{Complete schedule (in Portuguese): \url{https://www.gov.br/saude/pt-br/vacinacao/calendario}}. To do so, we construct a dictionary\footnote{{Available at \url{https://huggingface.co/gseovana/llama-vaccine-stance-ptbr-lora}}} that maps canonical vaccine names (e.g., \textit{BCG}, \textit{Poliomyelitis}, \textit{Yellow Fever}) to sets of associated lexical variants, including abbreviations, brand names, and colloquial references (for instance, mapping ``vacina da gotinha'' to Poliomyelitis). This dictionary is built from official documentation from the Ministry of Health and DataSUS\footnote{\url{https://www.gov.br/saude/pt-br/vacinacao/calendario}}, complemented by manual inspection of frequent n-grams in the corpus. Using this lexicon, we scan comment texts with regular expressions and assign each comment to one or more vaccines when a match is found. Once the mapping is established, we compute, for each vaccine and year, the number and proportion of comments in each stance category, enabling cross-vaccine comparisons that are consistent in both terminology and temporal coverage. Additionally, for some comparative visualizations involving vaccines with highly disparate frequencies, we apply z-score normalization to enhance interpretability and facilitate the comparison of temporal trends across vaccines\footnote{Z-score normalization transforms a value $x$ into $(x - \mu)/\sigma$, where $\mu$ is the mean and $\sigma$ is the standard deviation of the series. This scales different time series to a comparable range by centering them at zero and expressing deviations in units of standard deviation.}.

In addition to stance frequencies, we analyze the structure of antagonistic interactions by exploiting the reply tree available in YouTube comments. For each parent comment, we record its predicted stance and the stances of all direct replies. This allows us to estimate conditional reply probabilities of the form $P(s_{\text{reply}} \mid s_{\text{parent}})$, that is, the probability that a reply has stance $s_{\text{reply}}$ given that the parent has stance $s_{\text{parent}}$. We compute these probabilities separately for each phase (pre-pandemic, pandemic, and post-pandemic), with particular emphasis on cross-stance interactions, e.g.\ (reply) comments labeled as \textit{Against} made in reply to (parent) comments labeled as \textit{Favorable}. These measures provide a quantitative basis for assessing whether antagonistic responses intensify after the onset of the pandemic and how such changes vary over time and across different parts of the vaccination debate.

\subsection{Characterization of Dissemination Sources (RQ3)}

To address RQ3, we investigate how different segments of the YouTube ecosystem contribute to the dissemination and amplification of vaccine-related narratives. Our analysis focuses on characterizing channels according to their credibility and examining how this categorization relates to stance distributions and engagement patterns. To do that, we first classify each channel in the dataset into one of two categories: \textit{certified} or \textit{non-certified} information source. Following prior efforts \cite{Fonseca:2024}, a channel is labeled as \textit{certified} when it belongs to, or is formally affiliated with, an organization recognized by the \textit{Associação Nacional de Jornais} (ANJ), which establishes industry standards for journalistic ethics, editorial independence, and accountability in Brazil\footnote{\url{https://www.anj.org.br/associados/}}. To construct this classification, we match channel names and associated domains against the official ANJ membership registry and publicly available records of affiliated outlets.

Channels for which ANJ certification does not apply are labeled as \textit{non-certified}. 
Within this non-certified group, channels may fall into distinct subtypes: science communication, legacy media not affiliated to ANJ (e.g., international news organizations, such as BBC News and CNN), digital-native commentary. Regardless of subtype, all are treated uniformly as non-certified for the purposes of our analysis of source credibility and information flow.

Once channels are categorized, we compute a series of stance and engagement-driven metrics for each channel. For stance, we calculate the total number and relative proportion of comments in each class (\textit{Favorable}, \textit{Against}, \textit{Inconclusive}) associated with that channel's videos. For engagement, we aggregate likes and replies at the comment level, deriving per-channel statistics such as average likes per comment, average replies per comment, and the total number of unique users participating in discussions on that channel. We further derive polarization-oriented metrics, including the fraction of comments that are polarized (i.e., \textit{Favorable} or \textit{Against}) and the conditional probability that a comment on a given channel receives a reply of opposing stance. These comparative analyses provide the methodological basis for assessing how the structural properties of the YouTube channel ecosystem shape the dissemination and amplification of vaccination-related discourse, which we detail in the Results section.

\section{Results}
\label{sec:results}

This section reports the results aligned with our three RQ's. 

\begin{figure*}[!ht]
    \Description{A horizontal set of four bar charts comparing a Baseline model in blue and a Low Entropy model in gold. 
    The first three charts show Precision, Recall, and F1-Score for three categories: Against, Favorable, and Inconclusive. 
    In almost all categories, the gold bars (Low Entropy) are higher than the blue bars (Baseline). 
    Specifically, for the Favorable class, the Baseline has a significant performance drop with large error bars, while the Low Entropy model maintains high performance and stability. 
    The fourth chart on the right shows Overall Accuracy, where the Low Entropy model (gold) reaches approximately 0.9, clearly outperforming the Baseline which is at 0.85. 
    All charts include black vertical error bars indicating 95 percent confidence intervals.}
    \centering
    \includegraphics[width=0.9\textwidth]{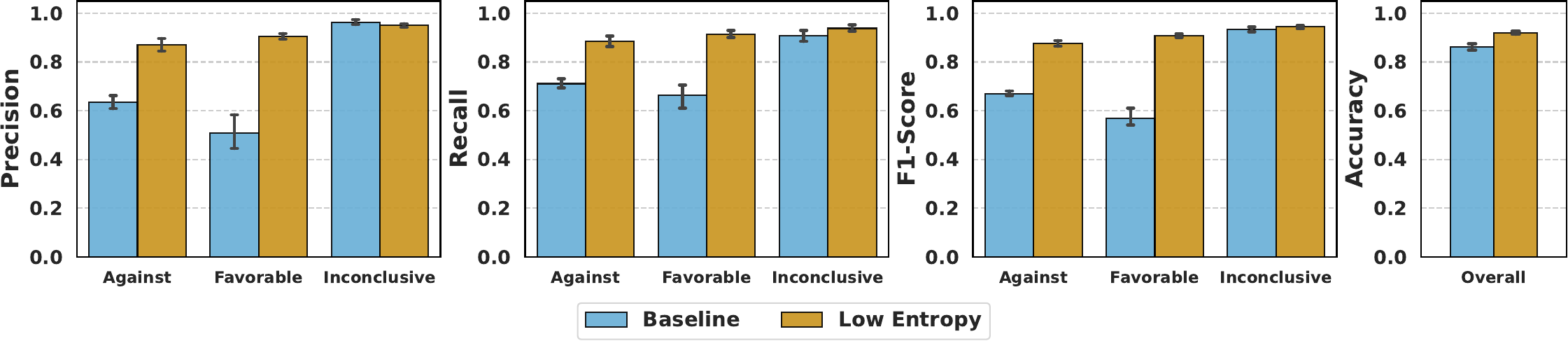}
    \caption{Comparison between the final baseline and low-entropy models with 95\% confidence intervals.}
    \label{fig:models-comparison}
\end{figure*}
\subsection{Stance Classification Models}
\label{subsec:models}

To address RQ1, we evaluate whether a semi-supervised stance classification pipeline can outperform a purely supervised baseline when applied to large-scale, imbalanced social media data. This section presents the performance of the supervised baseline model, the enrichment of the labeled dataset via low-entropy self-training, and the resulting improvements in the final classifier. Figure~\ref{fig:models-comparison} summarizes the average performance and 95\% confidence intervals for both models across folds, serving as the main visual reference for this discussion.

We begin with a fully supervised baseline model trained on the manually annotated dataset. As shown in Figure~\ref{fig:models-comparison}, the model exhibits stable but clearly constrained performance across the cross-validation folds, reflecting the strong class imbalance in the training data.  The majority \textit{Inconclusive} class is modeled reliably, achieving high precision ($0.96 \pm 0.02$), recall ($0.91 \pm 0.04$), and F1-score ($0.93 \pm 0.02$). In contrast, the minority classes show markedly weaker and, often times, less stable performance. The \textit{Against} class reaches a precision of $0.63 \pm 0.04$, recall of $0.71 \pm 0.03$, and F1-score of $0.67 \pm 0.02$, while the \textit{Favorable} class performs worst, with precision of $0.51 \pm 0.11$, recall of $0.66 \pm 0.08$, and F1-score of $0.57 \pm 0.06$. These limitations motivate the adoption of a semi-supervised enrichment strategy to stabilize learning and improve minority-class discrimination.

To mitigate these limitations, we apply the low-entropy semi-supervised sampling procedure described in Section~\ref{sec:methodology}. First, the baseline classifier generates class-probability distributions for the entire unlabeled dataset, from which we compute the Shannon entropy of each prediction. For each stance class, we then select a class-specific number of pseudo-labeled samples $k_c$, defined proportionally to the inverse frequency of that class in the labeled dataset, so as to partially correct for class imbalance. Within each class, we retain the $k_c$ samples with the lowest prediction entropy, corresponding to the most confident model outputs. This selection implicitly induces a class-specific entropy threshold $H(\hat{y}) < \epsilon_c$, defined as the maximum entropy value among the retained samples for that class. In practice, the resulting thresholds were $\epsilon_c = 0.0002$ for \textit{Against}, $\epsilon_c = 0.0041$ for \textit{Favorable}, and $\epsilon_c = 0.00004$ for \textit{Inconclusive}, corresponding to minimum class-probability scores of 1.0000, 0.9995, and 1.0000, respectively. These thresholds retain approximately the top 0.01\%, 1.13\%, and 0.002\% most confident predictions per class.

To counteract the dominance of the \textit{Inconclusive} class, we apply inverse-frequency sampling so that high-confidence \textit{Favorable} and \textit{Against} samples are preferentially included. The resulting pseudo-labeled dataset contains 2,004 comments (1,133 Favorable, 749 Against, and 122 Inconclusive). When combined with the original 3,476 manually labeled samples, we obtain a final training set of 5,480 examples (1,428 Favorable, 1,195 Against, and 2,857 Inconclusive), representing an increase of approximately 57.65\%. 
As expected, the enriched dataset substantially reduces class imbalance and introduces greater lexical diversity and contextual variability for minority classes.

Training the model on the enriched dataset yields consistently stronger performance across all classes (Figure~\ref{fig:models-comparison}), with gains are observed simultaneously in precision, recall, and F1-score. For the \textit{Against} class, precision increases from $0.63 \pm 0.04$ to $0.87 \pm 0.04$, recall from $0.71 \pm 0.03$ to $0.89 \pm 0.04$, and F1-score from $0.67 \pm 0.02$ to $0.88 \pm 0.02$. The \textit{Favorable} class, previously the most challenging, shows the largest improvement, with precision rising from $0.51 \pm 0.11$ to $0.91 \pm 0.02$, recall from $0.66 \pm 0.08$ to $0.91 \pm 0.03$, and F1-score from $0.57 \pm 0.06$ to $0.91 \pm 0.01$. 

Performance on the majority \textit{Inconclusive} class also improves, albeit more modestly, with precision increasing from $0.96 \pm 0.02$ to $0.95 \pm 0.01$, recall from $0.91 \pm 0.04$ to $0.94 \pm 0.02$, and F1-score from $0.93 \pm 0.02$ to $0.94 \pm 0.01$. Overall, Figure~\ref{fig:models-comparison} shows uniform upward shifts across all classes, accompanied by narrower confidence intervals, indicating reduced variance and more stable performance across folds.  Based on these results, we adopt the low-entropy model as the final classifier for large-scale stance inference. Its substantially higher performance, significantly improved performance on minority classes, and reduced variance across folds make it the most reliable model for all subsequent analyses.

\subsection{RQ2 – Temporal Dynamics and Polarization of Vaccination Discourse}

To address RQ2, we apply the final stance model to the full set of valid vaccination-related comments on YouTube. This large-scale inference step enables a systematic analysis of how volume, stance distribution, and polarization evolve over time and across vaccines in the Brazilian National Immunization Program.

Table~\ref{tab:stance-engagement-distribution} reports the final distribution of comments across the three stance categories, \textit{Favorable}, \textit{Against}, and \textit{Inconclusive}, as well as their associated engagement statistics (likes, replies, and unique participating users. The \textit{Inconclusive} class dominates the corpus, with 1,039,538 comments (74.4\% of all instances). This indicates that most interactions do not express explicit pro- or anti-vaccine positions, reflecting a heterogeneous space where neutral, informational, or off-topic exchanges are common. Rather than a purely polarized environment, the vaccination debate on YouTube appears as a heterogeneous space where ambiguity and non-committal discourse are the norm. 

Among polarized stances, \textit{Against} comments (204,179) outnumber \textit{Favorable} ones (152,940). Although the aggregate difference is modest, it becomes more salient when combined with engagement patterns. Overall, the stance distribution suggests that while explicit polarization is not the majority behavior, anti-vaccine narratives occupy a sizeable and persistent portion of the visible debate.

\begin{table}[t] 
\small 
\centering
\setlength{\tabcolsep}{4pt} 
\caption{Distribution of comments and engagement by stance.}
\label{tab:stance-engagement-distribution}
\begin{tabular}{lrrrr}
\toprule 
Stance       & \# Comments & \# Likes   & \# Replies & \# Users \\ \midrule
Favorable    & 152,940     & 598,780    & 66,959     & 91,633   \\
Against      & 204,179     & 838,977    & 79,137     & 115,037  \\
Inconclusive & 1,039,538   & 3,936,064  & 475,114    & 487,778  \\ \bottomrule
\end{tabular}
\vspace{-10pt}
\end{table}

On average, \textit{Against} comments receive slightly more likes per comment (4.1) than \textit{Favorable} ones (3.9), suggesting that skeptical or critical messages tend to attract a higher level of passive validation. On YouTube, likes act as lightweight yet visible markers of approval, potentially reinforcing perceptions of legitimacy and amplifying the perceived consensus. This pattern may increase the social impact of anti-vaccine content by making it appear more widely endorsed than it actually is. 

In contrast, \textit{Favorable} comments elicit more replies, with an average of 0.44 replies per comment compared to 0.36 for the \textit{Against} class. This suggests that pro-vaccine content more frequently triggers conversational engagement, including clarification requests and contestation, whereas anti-vaccine comments more often function as one-way statements that accumulate approval but receive fewer threaded responses. Overall, pro-vaccine discourse appears more dialogical, while anti-vaccine messages circulate as short, assertive statements requiring minimal interaction.

\subsubsection{Temporal and Vaccine-Level Dynamics}

\begin{figure*}[t]
\Description{Two side-by-side heatmaps showing vaccine mentions from 2018 to 2024. 
The left heatmap (a) is in green scales and represents Pro-vaccine mentions. 
The right heatmap (b) is in red scales and represents Anti-vaccine mentions. 
The vertical axis lists 17 different vaccines, including BCG, COVID-19, HPV, and MMR. 
The horizontal axis represents the years. 
The color intensity indicates the Z-Score normalized volume, ranging from negative 1.0 (lightest) to positive 1.0 (darkest). 
In both charts, there is a clear concentration of darker cells (higher volume) starting in 2020 and 2021, especially for the COVID-19 vaccine, which shows peak intensity across both pro and anti sentiments during the pandemic years. 
Some vaccines like Yellow Fever show a specific peak in 2018, while others like HPV and MMR show more distributed mentions over the years.}
\centering
    \begin{subfigure}{0.4\textwidth}
        \includegraphics[width=\textwidth]{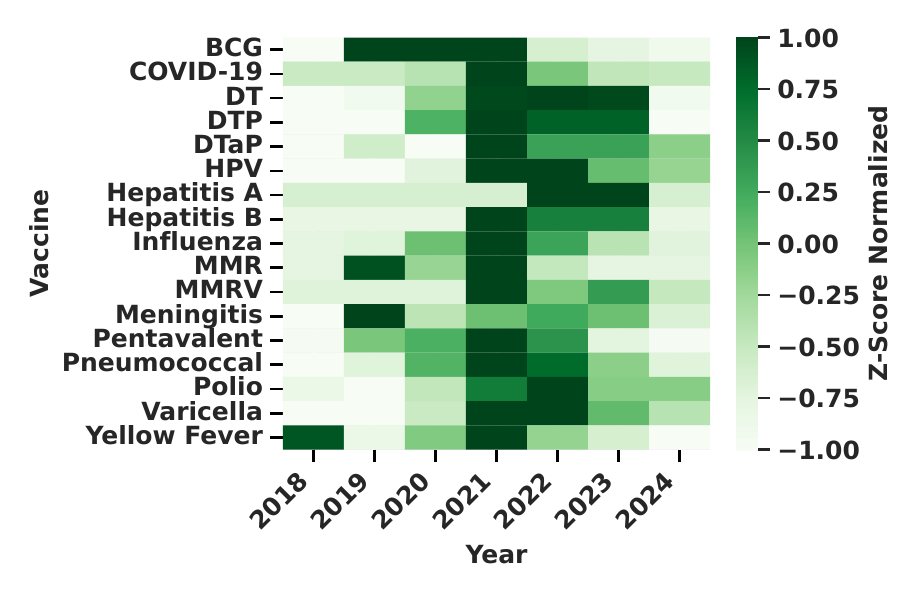}
        \caption{Pro-vaccine mentions by year.}
        \label{subfig:mencoes-favoraveis-por-ano-linha}
    \end{subfigure}
    \begin{subfigure}{0.4\textwidth}
        \includegraphics[width=\textwidth]{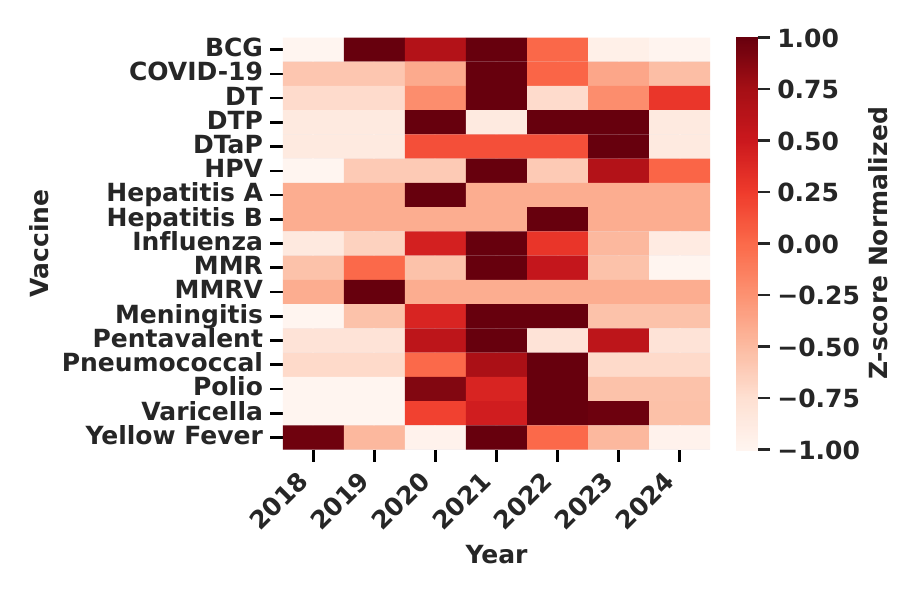}
        \vspace{-15pt}
        \caption{Anti-vaccine mentions by year.}
        \label{subfig:mencoes-contrarias-por-ano-linha}
    \end{subfigure}
    \caption{Distribution of vaccine-specific mentions over the years (z-score normalized per vaccine).} 
    \label{fig:heatmap-combined}
\end{figure*}

To examine how these dynamics unfold over time and across vaccines, we analyze the temporal distribution of mentions associated with each immunization in the Brazilian National Immunization Schedule. Figure~\ref{fig:heatmap-combined} shows the evolution of \textit{Favorable} and \textit{Against} vaccine mentions from 2018 to 2024. For each vaccine, we normalize values using z-score, which highlights years with anomalously high or low activity relative to that vaccine’s historical baseline. This row-wise normalization removes the bias of absolute volume, allowing us to compare temporal trends across different vaccines regardless of their overall popularity. 

The heatmaps indicate that vaccination discourse on YouTube is highly reactive to public health events. Yellow fever provides a clear example such that both favorable and opposing mentions peak sharply in 2018, coinciding with the major national outbreak.\footnote{\url{https://www12.senado.leg.br/noticias/especiais/especial-cidadania/surto-de-febre-amarela-desafia-vigilancia-de-epidemias/surto-de-febre-amarela-desafia-vigilancia-de-epidemias}}
From 2020 onward, the COVID-19 pandemic becomes a pivotal turning point. Nearly all vaccines exhibit marked shifts in their mention patterns, reflecting heightened public scrutiny of vaccination as a whole. Influenza vaccines also gain visibility during this period, with both supportive and critical mentions increasing after 2020, likely driven by frequent comparisons among influenza, the common cold, and COVID-19.\footnote{\url{https://www.ncoa.org/article/whats-the-difference-between-flu-and-covid-a-guide-for-older-adults/}} This suggests that respiratory-related vaccines occupy a central role in public attention during the pandemic.

Notably, several vaccines display simultaneous increases in both pro- and anti-vaccine mentions, particularly in 2021. For example, DTP, meningitis, pentavalent, Measles, Mumps, Rubella, Varicella (MMRV), and Measles, Mumps, and Rubella (MMR) show elevated activity in both heatmaps. Such parallel growth across opposing stances is not merely a sign of increased discourse volume, but indicates an intensification of polarization, in which heightened visibility of vaccination stimulates competing narratives and more direct contestation between supportive and skeptical positions. This dual escalation highlights how crisis periods amplify not only engagement but also ideological divergence.

Row-wise normalization further shows that crises act as focal points for attention concentration. In 2018, yellow fever effectively monopolizes the debate, whereas in 2019 the focus diversifies. On the favorable side,  Bacillus Calmette-Guérin (BCG), MMR, and meningitis become more prominent. On the opposing side, BCG and hepatitis B stand out. In 2020, the onset of the COVID-19 emergency triggers a broad expansion of anti-vaccine attention across multiple immunizations. In 2021, the picture becomes more complex. While several routine vaccines attract negative attention, nearly all vaccines also experience an increase in favorable discourse. This shift is plausibly linked to the rollout of COVID-19 vaccination in Brazil,\footnote{\url{https://www.cnnbrasil.com.br/nacional/primeira-pessoa-e-vacinada-contra-covid-19-no-brasil/}} which renewed public interest in the historical and contemporary importance of immunization.

A closer examination of the heatmaps also reveals substantial heterogeneity across vaccines. Some vaccines, such as BCG and HepA, exhibit comparatively stable temporal profiles, with fluctuations that remain closer to their historical baselines. In contrast, others, such as influenza, varicella, pentavalent, and especially COVID-19, display highly volatile trajectories, with large annual swings that reflect their sensitivity to news cycles, outbreaks, and shifting public concerns. This structural heterogeneity suggests that not all vaccines occupy equivalent positions in public discourse where some are consistently salient due to their epidemiological relevance, while others become prominent only when activated by specific events or controversies.

In the subsequent years (2022–2023), attention becomes more fragmented, with different vaccines intermittently emerging in both favorable and opposing narratives. This pattern suggests a residual discursive effect of the pandemic, that is, sustained interest in vaccination persists, but it is distributed across a wider set of immunizations rather than concentrated on a single crisis. The apparent decline in 2024 is largely explained by the truncated collection window (first 7 months only). 

Taken together, these results reveal a highly dynamic discourse ecosystem that responds directly to epidemiological and media events. During crises, public attention tends to concentrate on a small subset of vaccines, while in calmer periods the debate diffuses across the broader immunization schedule. The simultaneous rise of pro and anti-vaccine mentions during certain years further underscores the role of crisis conditions in amplifying polarization, not only engagement. This reinforces the central role of public health dynamics in shaping how vaccination is discussed and contested online.

\subsubsection{Polarization and Antagonistic Interaction}

\begin{figure}[!t]
\Description{A set of three side-by-side bar charts titled Reply Probability. 
Each chart shows the probability of a reply being Against (red) or Favorable (green) based on the Parent Comment Class (Against or Favorable) on the horizontal axis. 
In the first chart, there is a clear echo effect: Against parents have a 0.59 probability of Against replies, while Favorable parents have a 0.69 probability of Favorable replies. 
The second chart shows a more balanced distribution, with probabilities hovering around 0.43 to 0.57 for both classes. 
The third chart shows a different trend where Against parent comments receive Against replies with 0.67 probability, but Favorable parents receive an equal split of 0.50 for both Against and Favorable replies. 
The vertical axis ranges from 0.0 to 1.0, and exact probability values are printed above each bar.}
\centering
    \begin{subfigure}{0.38\columnwidth}
        \includegraphics[width=\textwidth]{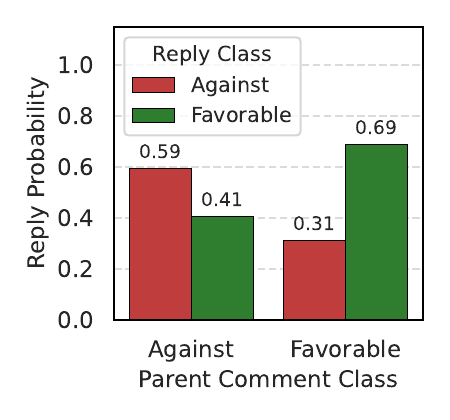}
        \caption{Pre-pandemic}
        \label{subfig:pre-pandemic}
    \end{subfigure}
    \hfill 
    \begin{subfigure}{0.3\columnwidth}
        \includegraphics[trim={47 0 0 0},clip,width=\textwidth]{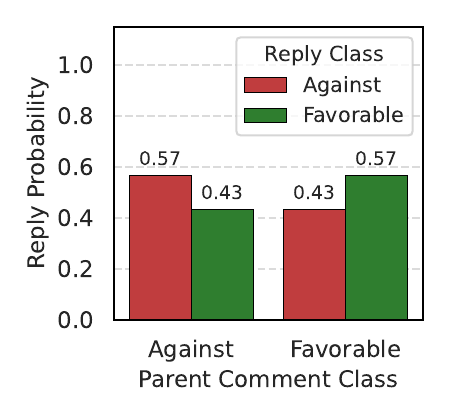}
        \caption{During pandemic}
        \label{subfig:during-pandemic}
    \end{subfigure}
    \hfill
    \begin{subfigure}{0.3\columnwidth}
        \includegraphics[trim={47 0 0 0},clip,width=\textwidth]{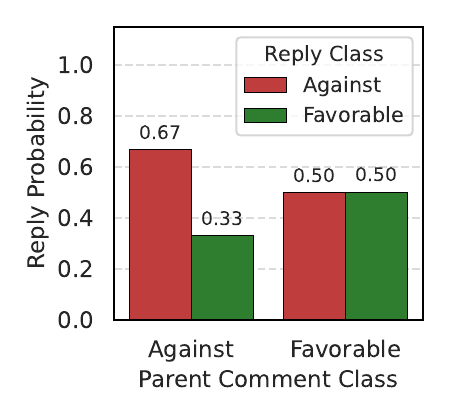}
        \caption{Post-pandemic}
        \label{subfig:post-pandemic}
    \end{subfigure}
    \caption{Conditional reply probabilities by stance in pre, during, and post-pandemic periods.}
    \label{fig:conditinal-reply-probability}
\end{figure}

To capture the intensity of polarization in user interactions, we analyze conditional reply probabilities between stance categories using the reply-tree structure available in YouTube comments. Figure~\ref{fig:conditinal-reply-probability} reports conditional reply probabilities of the form $P(s_{\text{reply}} \mid s_{\text{parent}})$ across three periods (pre-pandemic, during the pandemic, and post-pandemic), allowing us to examine how antagonistic and supportive interactions evolve over time. We focus in particular on \textit{cross-stance replies}, such as \textit{Against} replies to \textit{Favorable} parents and \textit{Favorable} replies to \textit{Against} parents, which indicate contestational or rebuttal dynamics.

A first pattern emerges in the pre-pandemic period (Figure~\ref{subfig:pre-pandemic}). Here, the pro-vaccine community exhibits high internal cohesion, with \textit{Favorable} replies to \textit{Favorable} comments reaching a probability of $0.69$, the highest level of in-group alignment observed across periods. In terms of cross-stance engagement, an asymmetry appears: \textit{Favorable} replies to \textit{Against} parents occur with probability $0.41$, whereas \textit{Against} replies to \textit{Favorable} parents occur with probability $0.31$. This suggests that, initially, cross-stance interaction was more often driven by pro-vaccine users, consistent with corrective or debunking behavior, while the skeptical group was comparatively less likely to engage pro-vaccine comments.

During the pandemic period (Figure~\ref{subfig:during-pandemic}), reply patterns shift towards structural symmetry. In-group alignment decreases to $0.57$ for both stances, and cross-stance replies become perfectly balanced: $P(\textit{Against}\mid\textit{Favorable}) = 0.43$ and $P(\textit{Favorable}\mid\textit{Against}) = 0.43$. This indicates that, at the height of the crisis, the discussion becomes a more symmetric arena of contestation. Consistent with this interactional shift, Table~\ref{tab:polarized-distribution} shows that the polarized proportion nearly doubles from 13.62\% (pre-pandemic) to 26.02\% (during pandemic), reflecting a substantial expansion of cross-stance confrontation.

A more pronounced reconfiguration appears in the post-pandemic period (Figure~\ref{subfig:post-pandemic}). \textit{Against} replies to \textit{Against} parents rise to $0.67$, indicating stronger in-group reinforcement. In parallel, the pro-vaccine side becomes less internally supported: both in-group support and cross-stance exposure converge at $0.50$, i.e., $P(\textit{Favorable}\mid\textit{Favorable}) = 0.50$ and $P(\textit{Against}\mid\textit{Favorable}) = 0.50$. This suggests a weakening of pro-vaccine conversational buffering, contrasting with the consolidation of skeptical alignment.

\begin{table}[t!]
\centering
\small
\caption{Distribution of polarized comments across pre, during, and post-pandemic periods.}
\label{tab:polarized-distribution}
\resizebox{0.8\columnwidth}{!}{%
\begin{tabular}{lccc}
\hline
\multirow{2}{*}{Period}          & Total of & \# Polarized & Polarized \\
& Comments & Comments & Proportion  \\\hline
Pre-pandemic    & 74,174            & 10,104                & 13.62\%               \\
During Pandemic & 1,219,820         & 317,406               & 26.02\%                \\
Post-pandemic   & 102,663           & 29,609                & 28.84\%                \\ \hline
\end{tabular}%
}
\end{table}

Taken together, the three periods reveal a clear temporal trajectory. Pre-pandemic interactions are characterized by pro-vaccine cohesion and asymmetric corrective engagement; the pandemic period is associated with a more symmetric pattern of contestation ($0.43$ vs.\ $0.43$); and the post-pandemic period shows increased skeptical reinforcement ($0.67$) alongside reduced pro-vaccine in-group support ($0.50$). Although platform ranking and recommendation mechanisms may influence which threads gain greater visibility, our analysis captures the structure of interactions as they unfold within the YouTube ecosystem. In this sense, the findings demonstrate that polarization is not merely distributional, but embedded in the \textit{interactional structure} of conversations, shaping how users confront, challenge, or reinforce vaccine-related narratives over time.

\vspace{-5pt}
\subsection{RQ3: Sources of Dissemination Across Stance Categories}\label{subsec:rq3}

To address RQ3, we investigate where polarized discourse is most concentrated on YouTube by identifying the channels and videos that accumulate the highest volumes of favorable, opposing, and polarized comments. Table \ref{tab:channel_slopegraph} provides a comparative cross-ranking analysis of the leading channels involved in the vaccination debate ranked by total comment volume in descending order. The table is structured with the channel names in the center column, using arrows to map their rank and volume of pro-vaccine comments on the left versus anti-vaccine comments on the right. To contextualize these patterns, we integrate two dimensions: (1) channel association with a professionally certified news outlet, and (2) channel classification into \textit{Legacy News Media (LNM)}, \textit{Science/Health Communicators (SC)}, or \textit{Digital-Native Commentary (DC)}. Thus, our focus is to assess whether structured polarization emerges primarily through the professional news ecosystem, scientific outreach channels, or unregulated/independent digital commentary spaces.

\textbf{Channel-Level Dynamics}. Three structural findings emerge from this visualization. First, non-certified channels are the primary arenas for both supportive and opposing vaccination discourse, reflecting the hybrid and decentralized nature of public health communication on YouTube. This is quantitatively evident, as 82.4\% of all anti-vaccine comments occur on channels that do not belong to the ANJ-certified media ecosystem. The pronounced overlap between the two rankings -- where 14 of the top 15 channels appear in both lists -- confirms that these independent spaces function as the central battlegrounds for the debate.

Second, science communicators occupy a central influential position as they anchor the most active discussions and attract diverse forms of stance expression. Channels such as \textit{Julio Pereira – Neurocirurgião} and \textit{Olá, Ciência!} rank highest on both sides, indicating that they function not merely as channels for science dissemination, but as arenas where scientific authority is actively contested by a skeptical audience.

Third, legacy media, while present, act more as agenda amplifiers than as the primary loci of polarized engagement. While outlets like \textit{BBC News Brasil} and \textit{CNN Brasil} appear in the rankings, the volume of opposition they receive exceeds their supportive engagement; yet, they fail to surpass the interaction levels of top independent creators. This pattern suggests that although news media set the agenda regarding vaccination, science communication channels host the most intensive discussions, generating both support and contestation.

\begin{table}[tbp!]
\centering
\small
\caption{Cross-ranking Analysis: ANJ certification indicated (\cmark{} Yes, \xmark{} No, -- N/A). Ch. Type key: SC=Science Communicator, LNM=Legacy News Media, DC=Digital-native Commentary. Channels are sorted by total comments.}
\label{tab:channel_slopegraph}
\scriptsize
\setlength{\tabcolsep}{3pt}
\renewcommand{\arraystretch}{1.4}

\tikzset{
    link/.style={
        lineColor,
        thick,
        -{Stealth[length=2mm, width=1.2mm]},
        opacity=0.6,
        shorten <= 2pt,
        shorten >= 2pt
    },
}

\newcolumntype{S}{p{0.4cm}}

\begin{tabular}{c c c r S c S l c}
\toprule
\textbf{Channel} & \multirow{2}{*}{\textbf{ANJ}} & \multicolumn{2}{c}{\textbf{\textcolor{proColor}{Pro-Vaccine}}} & & \multirow{2}{*}{\textbf{Channel Name}} & & \multicolumn{2}{c}{\textbf{\textcolor{antiColor}{Anti-Vaccine}}} \\

\textbf{Type} & & \textbf{Rk} & \textbf{Count} & & & & \textbf{Count} & \textbf{Rk} \\
\midrule


SC & -- & 1 & \tikzmarknode{pro1}{29,345} & &
\tikzmarknode{cen1}{Julio Pereira – Neurosurgeon} & &
\tikzmarknode{anti1}{27,121} & 1 \\

SC & -- & 2 & \tikzmarknode{pro2}{19,487} & &
\tikzmarknode{cen2}{Hello, Science! (Olá, Ciência!)} & &
\tikzmarknode{anti2}{21,117} & 2 \\

LNM & -- & 3 & \tikzmarknode{pro3}{12,569} & &
\tikzmarknode{cen3}{BBC News Brasil} & &
\tikzmarknode{anti3}{17,934} & 3 \\

LNM & \cmark & 4 & \tikzmarknode{pro4}{9,451} & &
\tikzmarknode{cen4}{UOL} & &
\tikzmarknode{anti4}{14,052} & 4 \\

LNM & \cmark & 5 & \tikzmarknode{pro5}{5,740} & &
\tikzmarknode{cen5}{TV Cultura Journalism} & &
\tikzmarknode{anti5}{10,117} & 5 \\

LNM & \xmark & 6 & \tikzmarknode{pro6}{5,589} & &
\tikzmarknode{cen6}{Band Journalism} & &
\tikzmarknode{anti6}{9,059} & 6 \\

LNM & \xmark & 7 & \tikzmarknode{pro7}{3,988} & &
\tikzmarknode{cen7}{SBT News} & &
\tikzmarknode{anti7}{8,962} & 7 \\

LNM & -- & 8 & \tikzmarknode{pro8}{3,866} & &
\tikzmarknode{cen8}{CNN Brasil} & &
\tikzmarknode{anti8}{5,916} & 8 \\

LNM & \xmark & 9 & \tikzmarknode{pro9}{2,359} & &
\tikzmarknode{cen9}{Jovem Pan News} & &
\tikzmarknode{anti9}{5,716} & 9 \\

SC & -- & 10 & \tikzmarknode{pro10}{2,290} & &
\tikzmarknode{cen10}{Prof. Lysandro Borges} & &
\tikzmarknode{anti10}{5,280} & 10 \\

DC & \xmark & 11 & \tikzmarknode{pro11}{1,822} & &
\tikzmarknode{cen11}{Veja.com} & &
\tikzmarknode{anti11}{3,062} & 11 \\

LNM & \xmark & 12 & \tikzmarknode{pro12}{1,762} & &
\tikzmarknode{cen12}{Record News} & &
\tikzmarknode{anti12}{2,632} & 12 \\

DC & \xmark & 13 & \tikzmarknode{pro13}{1,638} & &
\tikzmarknode{cen13}{Spotniks} & &
\tikzmarknode{anti13}{2,387} & 13 \\

DC & \xmark & 14 & \tikzmarknode{pro14}{1,551} & &
\tikzmarknode{cen14}{Unknown Facts} & &
\tikzmarknode{anti14}{2,219} & 14 \\

LNM & -- & 15 & \tikzmarknode{pro15}{1,506} & &
\tikzmarknode{cen15}{\textbf{TV Brasil}} & &
\tikzmarknode{anti15}{1,981} & 15 \\

SC & -- &  & & &
\tikzmarknode{cen16}{\textbf{Átila Iamarino}} & &
 &  \\

\bottomrule
\end{tabular}

\begin{tikzpicture}[overlay, remember picture]
    \draw[link] (cen1.west) -- (pro1.east);
    \draw[link] (cen2.west) -- (pro2.east);
    \draw[link] (cen3.west) -- (pro3.east);
    \draw[link] (cen4.west) -- (pro4.east);
    \draw[link] (cen5.west) -- (pro6.east); 
    \draw[link] (cen6.west) -- (pro5.east); 
    \draw[link] (cen7.west) -- (pro7.east);
    \draw[link] (cen8.west) -- (pro8.east);
    \draw[link] (cen9.west) -- (pro10.east); 
    \draw[link] (cen10.west) -- (pro9.east); 
    \draw[link] (cen11.west) -- (pro12.east);
    \draw[link] (cen12.west) -- (pro15.east);
    \draw[link] (cen13.west) -- (pro11.east);
    \draw[link] (cen14.west) -- (pro13.east);
    \draw[link] (cen16.west) -- (pro14.east); 

    \draw[link] (cen1.east) -- (anti1.west);
    \draw[link] (cen2.east) -- (anti2.west);
    \draw[link] (cen3.east) -- (anti3.west);
    \draw[link] (cen4.east) -- (anti4.west);
    \draw[link] (cen5.east) -- (anti5.west);
    \draw[link] (cen6.east) -- (anti7.west);
    \draw[link] (cen7.east) -- (anti6.west);
    \draw[link] (cen8.east) -- (anti10.west);
    \draw[link] (cen9.east) -- (anti8.west);
    \draw[link] (cen10.east) -- (anti9.west);
    \draw[link] (cen11.east) -- (anti11.west);
    \draw[link] (cen12.east) -- (anti12.west);
    \draw[link] (cen13.east) -- (anti14.west);
    \draw[link] (cen14.east) -- (anti13.west);
    \draw[link] (cen15.east) -- (anti15.west);
\end{tikzpicture}
\end{table}

\begin{table*}[!t]
\centering
\caption{Top 15 videos ranked by different counts. (Top) Largest number of anti-vaccine comments; (Middle) Largest number of pro-vaccine comments; (Bottom) Largest number of polarized comments, i.e., opposing + favorable.  Titles were translated to English.}
\label{tab:top15}
\scriptsize
\begin{tabular}{clcccc}
\multicolumn{6}{c}{\textbf{Ranked by counts of anti-vaccine comments}}\\ \hline
Rank & Video Title (Translated) & Video Link & \# Anti Comments & \# Views & \# Likes \\ \hline
1 & Covid-19: what might happen to those who reject the vaccine? &
\href{https://www.youtube.com/watch?v=bfbyImPA938}{Watch} &
3,796 & 509,383 & 23,922 \\
2 & Three diseases that are after-effects of Covid and nobody talks about! &
\href{https://www.youtube.com/watch?v=Hu4jz-pPlcQ}{Watch} &
3,109 & 2,528,525 & 138,783 \\
3 & Covid vaccine: does AstraZeneca cause thrombosis? &
\href{https://www.youtube.com/watch?v=jAd-GxRdBbY}{Watch} &
3,027 & 2,117,694 & 100,382 \\
4 & Health professionals dismiss the link between a teenager's death and vaccination &
\href{https://www.youtube.com/watch?v=eoCKbrRjLc4}{Watch} &
2,667 & 285,228 & 15,687 \\
5 & The search for people who may never catch Covid &
\href{https://www.youtube.com/watch?v=07iN9CqWh_Q}{Watch} &
2,308 & 3,120,004 & 177,076 \\
6 & He suspects Covid vaccines. She is a Butantan scientist. We put them to talk. &
\href{https://www.youtube.com/watch?v=L24UIaHb_8A}{Watch} &
2,219 & 1,386,486 & 80,386 \\
7 & Covid vaccine: what they did not tell you about the third dose &
\href{https://www.youtube.com/watch?v=jrSNYg6ngnY}{Watch} &
2,138 & 2,760,791 & 137,807 \\
8 & Four factors that make vaccinated people worsen from Covid (that they don't tell you) &
\href{https://www.youtube.com/watch?v=DBvEWHTxhb0}{Watch} &
2,137 & 793,289 & 50,451 \\
9 & Vaccinated woman dies after second dose &
\href{https://www.youtube.com/watch?v=3dngPfDT80g}{Watch} &
1,991 & 642,203 & 16,736 \\
10 & Cases emerge of people who died of Covid-19 after vaccination &
\href{https://www.youtube.com/watch?v=f1DzS70LJPY}{Watch} &
1,889 & 563,205 & 18,385 \\
11 & Why is there no HIV vaccine yet? (And Covid already has one) &
\href{https://www.youtube.com/watch?v=R3h5YMAAgIs}{Watch} &
1,887 & 3,176,542 & 152,073 \\
12 & Deaths after Covid-19 vaccination under investigation &
\href{https://www.youtube.com/watch?v=VgwaYg6Nfa4}{Watch} &
1,765 & 409,363 & 11,591 \\
13 & Young woman suffers thrombosis and dies; family suspects vaccine reaction &
\href{https://www.youtube.com/watch?v=C1vVwGeZvK4}{Watch} &
1,751 & 385,491 & 15,034 \\
14 & Now Lula forces Brazilians to take the Covid-19 vaccine... &
\href{https://www.youtube.com/watch?v=s8tT4LeBYy0}{Watch} &
1,733 & 171,778 & 13,354 \\
15 & Covid-19: “I will NOT take the vaccine!!!” &
\href{https://www.youtube.com/watch?v=hkp7wxV34c8}{Watch} &
1,706 & 379,518 & 16,838 \\ \hline
\end{tabular}%
\\ [3ex]
\begin{tabular}{clcccc}
\multicolumn{6}{c}{\textbf{Ranked by counts of pro-vaccine comments}}\\
\hline
Rank & Video Title (Translated) & Video Link & \# Pro Comments & \# Views & \# Likes \\ \hline
1 & Covid vaccine: AstraZeneca --- Everything you need to know \#1 &
\href{https://www.youtube.com/watch?v=PoqRJRZWpZs}{Watch} &
2,012 & 1,246,562 & 97,079 \\
2 & Covid vaccine: does AstraZeneca cause thrombosis? &
\href{https://www.youtube.com/watch?v=jAd-GxRdBbY}{Watch} &
1,989 & 2,117,694 & 100,382 \\
3 & He suspects Covid vaccines. She is a Butantan scientist. We put them to talk. &
\href{https://www.youtube.com/watch?v=L24UIaHb_8A}{Watch} &
1,822 & 1,386,486 & 80,386 \\
4 & Why does the BCG vaccine leave a mark on the arm only in Brazil? &
\href{https://www.youtube.com/watch?v=aFDiIUr2a-8}{Watch} &
1,531 & 2,016,589 & 170,228 \\
5 & Covid vaccine: Pfizer - Everything you need to know \#2 &
\href{https://www.youtube.com/watch?v=i58ZIFyuWac}{Watch} &
1,471 & 832,811 & 60,961 \\
6 & Covid-19: what might happen to those who reject the vaccine? &
\href{https://www.youtube.com/watch?v=bfbyImPA938}{Watch} &
1,429 & 509,383 & 23,922 \\
7 & Covid vaccine: what they did not tell you about the third dose &
\href{https://www.youtube.com/watch?v=jrSNYg6ngnY}{Watch} &
1,378 & 2,760,791 & 137,807 \\
8 & Covid-19: side effects of the Pfizer vaccine explained &
\href{https://www.youtube.com/watch?v=yZ1UDPO2Ak4}{Watch} &
1,118 & 440,481 & 12,783 \\
9 & Covid vaccine: will AstraZeneca cause deaths in the future? &
\href{https://www.youtube.com/watch?v=o0749cT5xBI}{Watch} &
1,094 & 652,312 & 23,464 \\
10 & The search for people who may never catch Covid &
\href{https://www.youtube.com/watch?v=07iN9CqWh_Q}{Watch} &
1,091 & 3,120,004 & 177,076 \\
11 & Three diseases that are after-effects of Covid and nobody talks about! &
\href{https://www.youtube.com/watch?v=Hu4jz-pPlcQ}{Watch} &
1,077 & 2,528,525 & 138,783 \\
12 & Covid vaccine: Janssen - Everything you need to know \#3 &
\href{https://www.youtube.com/watch?v=FZowUdBnEGA}{Watch} &
1,058 & 725,411 & 47,444 \\
13 & Four factors that make vaccinated people worsen from Covid (that they don't tell you) &
\href{https://www.youtube.com/watch?v=DBvEWHTxhb0}{Watch} &
1,006 & 793,289 & 50,451 \\
14 & AstraZeneca vaccine - Future deaths and complications? &
\href{https://www.youtube.com/watch?v=OqK2b1ehHqw}{Watch} &
977 & 1,518,676 & 44,640 \\
15 & Vaccinated woman dies after second dose &
\href{https://www.youtube.com/watch?v=3dngPfDT80g}{Watch} &
970 & 642,203 & 16,736 \\ \hline
\end{tabular}%
\\ [3ex]
\begin{tabular}{clcccc}
\multicolumn{6}{c}{\textbf{Ranked by counts of polarized comments}}\\
\hline
Rank & Video Title (Translated) & Video Link & \# Polarized Comments & \# Views & \# Likes \\ \hline
1 & Covid-19: what might happen to those who reject the vaccine? &
\href{https://www.youtube.com/watch?v=bfbyImPA938}{Watch} &
5,225 & 509,383 & 23,922 \\
2 & Covid vaccine: does AstraZeneca cause thrombosis? &
\href{https://www.youtube.com/watch?v=jAd-GxRdBbY}{Watch} &
5,016 & 2,117,694 & 100,382 \\
3 & Three diseases that are after-effects of Covid and nobody talks about! &
\href{https://www.youtube.com/watch?v=Hu4jz-pPlcQ}{Watch} &
4,186 & 2,528,525 & 138,783 \\
4 & He suspects Covid vaccines. She is a Butantan scientist. We put them to talk. &
\href{https://www.youtube.com/watch?v=L24UIaHb_8A}{Watch} &
4,041 & 1,386,486 & 80,386 \\
5 & Covid vaccine: what they did not tell you about the third dose &
\href{https://www.youtube.com/watch?v=jrSNYg6ngnY}{Watch} &
3,516 & 2,760,791 & 137,807 \\
6 & The search for people who may never catch Covid &
\href{https://www.youtube.com/watch?v=07iN9CqWh_Q}{Watch} &
3,399 & 3,120,004 & 177,076 \\
7 & Covid vaccine: AstraZeneca - Everything you need to know \#1 &
\href{https://www.youtube.com/watch?v=PoqRJRZWpZs}{Watch} &
3,268 & 1,246,562 & 97,079 \\
8 & Health professionals dismiss the link between a teenager's death and vaccination &
\href{https://www.youtube.com/watch?v=eoCKbrRjLc4}{Watch} &
3,207 & 285,228 & 15,687 \\
9 & Four factors that make vaccinated people worsen from Covid (that they don't tell you) &
\href{https://www.youtube.com/watch?v=DBvEWHTxhb0}{Watch} &
3,143 & 793,289 & 50,451 \\
10 & Why does the BCG vaccine leave a mark on the arm only in Brazil? &
\href{https://www.youtube.com/watch?v=aFDiIUr2a-8}{Watch} &
3,020 & 2,016,589 & 170,228 \\
11 & Vaccinated woman dies after second dose &
\href{https://www.youtube.com/watch?v=3dngPfDT80g}{Watch} &
2,961 & 642,203 & 16,736 \\
12 & Cases emerge of people who died of Covid-19 after being vaccinated &
\href{https://www.youtube.com/watch?v=f1DzS70LJPY}{Watch} &
2,788 & 563,205 & 18,385 \\
13 & Covid-19: “I will NOT take the vaccine!!!” &
\href{https://www.youtube.com/watch?v=hkp7wxV34c8}{Watch} &
2,660 & 379,518 & 16,838 \\
14 & Covid vaccine: will AstraZeneca cause deaths in the future? &
\href{https://www.youtube.com/watch?v=o0749cT5xBI}{Watch} &
2,653 & 652,312 & 23,464 \\
15 & Why is there no HIV vaccine yet? (And Covid already has one) &
\href{https://www.youtube.com/watch?v=R3h5YMAAgIs}{Watch} &
2,622 & 3,176,542 & 152,073 \\ \hline
\end{tabular}
\end{table*}

\textbf{Video-Level Concentration.} We now shift from channel-level aggregation to video-level concentration.
Table~\ref{tab:top15} (top) lists the videos with the highest number of anti-vaccine comments, with titles translated into English. The majority address controversies, health risks, or vaccine side effects, topics that historically tend to elicit hesitation or fear-based reactions. Several videos also relate to breaking news about adverse events or alleged complications, reinforcing how event-driven cycles shape online contestation.

Based on Table~\ref{tab:top15} (middle), which lists the videos with highest volume of favorable comments, we note that this type of engagement often concentrates in explanatory content and videos that address misconceptions. Meanwhile, videos with the highest total polarized comments (favorable + opposing) include either debunking material or controversy-oriented narratives, illustrating how polarization can emerge simultaneously through informational correction and sensational framing.

Table~\ref{tab:top15} (bottom) highlights the videos that concentrate the highest combined volume of favorable and opposing comments, offering a direct view of where explicit confrontation between pro- and anti-vaccine positions is most intense. A clear pattern emerges: the majority of highly polarized videos are not explicitly anti-vaccine, but instead address ambiguous, contested, or high-visibility topics such as adverse events, vaccine side effects, or hypothetical risks. This indicates that polarization is amplified not only by misinformative content, but also by informational or journalistic material that touches on uncertainty, controversy, or medical complexity.

A second pattern is the disproportionate presence of AstraZeneca-related videos. Five of the top fifteen most polarized videos focus on the AstraZeneca vaccine, suggesting that this immunization became a symbolic focal point of public controversy. The strong association between AstraZeneca and discussions of thrombosis or long-term effects appears to sustain both criticism and corrective responses, effectively transforming these videos into ``polarity magnets''. Another notable feature is the prominence of dialogical formats, such as videos that stage a conversation between a vaccine-skeptical individual and a medical expert. These formats attract intense engagement from both sides, consistent with literature showing that public debates, even when evidence-based, can invite contestational participation and strengthen identity-based reactions.

Finally, the degree of polarization is not simply a function of viewership. Several videos with relatively modest audience sizes generate high volumes of polarized comments, indicating that topic sensitivity plays a larger role than reach in triggering stance confrontation. In sum, polarization tends to emerge most strongly around content that evokes uncertainty, perceived controversy, or high emotional salience, rather than through overtly anti-vaccine messaging alone.

The combination of channel typology, ANJ certification status, and stance distribution reveals a three-tiered structure of dissemination. Certified legacy news outlets contribute substantially to the overall visibility of vaccine content, but they are not the primary sites where polarization concentrates. Instead, science communication channels, many of which are non-certified but highly trusted by their audiences,  emerge as the central hubs of engagement, attracting both strong support and intense contestation. In parallel, digital-native commentary channels, although generally smaller, play a disproportionate role in amplifying contentious narratives and generating high densities of opposing comments. Taken together, these patterns indicate that vaccine-related polarization on YouTube is shaped not by traditional journalism alone, but by a hybrid information ecosystem in which institutional media, scientific communicators, and informal commentary channels interact dynamically, each contributing in distinct ways to the circulation and contestation of vaccination narratives.

\section{Conclusions and Future Work}
\label{sec:conclusions}

Vaccination debates have become a central arena where public health, political polarization, and platform-mediated communication intersect. Using Brazil as a longitudinal case study, we examined how vaccine-related discourse evolved on YouTube across pre-pandemic, pandemic, and post-pandemic periods. In a country with a historically strong immunization program but recent institutional contestation, this trajectory offers insight into how public health polarization emerges, restructures, and persists on large-scale digital platforms.

Methodologically, this study shows that semi-supervised stance modeling with low-entropy self-training enables scalable analysis of highly imbalanced and ambiguous user-generated content. Confidence-based selection substantially improves minority-class performance while limiting manual annotation costs, offering a practical framework for longitudinal studies of polarized discourse in data-scarce contexts.

Substantively, our findings indicate that vaccine polarization is dynamic rather than static. During the pandemic, interactions converged into a more symmetric pattern of contestation, coinciding with a period of heightened uncertainty and intensified reciprocal engagement. In the post-pandemic period, polarization became increasingly asymmetric, marked by the consolidation of skeptical communities and weakened pro-vaccine cohesion. These shifts demonstrate that polarization is embedded not only in stance distributions but in the interactional structure of conversations.

At the ecosystem level, polarization emerged primarily within a hybrid media environment dominated by science communicators and digital-native channels rather than legacy journalism. These channels function simultaneously as spaces of trust-building and sites of epistemic conflict, revealing structural tensions in contemporary health communication. From a sociotechnical perspective, the results suggest that interventions focused solely on content moderation may be insufficient, as polarization is also shaped by interactional dynamics and evolving trust structures.

This study has limitations. The analysis is restricted to YouTube and captures expressed stances rather than passive exposure or offline effects. As an observational design, it does not establish causality and may reflect platform-level influences such as recommendation systems. Methodologically, model confidence remains a proxy for accuracy. A systematic manual audit would help assess potential semantic drift and further validate the self-training procedure. Future research may also examine the internal structure of the Inconclusive class and explore how specific vaccines are differentially framed across groups. Finally, the framework can be extended to cross-platform and decentralized ecosystems \cite{nobre2022more}, integrated with community detection approaches \cite{santos2025high, oliveira2024framework, oliveira2025network}, enriched with multimodal signals, and linked to vaccination uptake or epidemiological outcomes.

\appendix
\section{Detailed Search Strategy}
\label{appendix:search_strategy}

To ensure comprehensive coverage of the Brazilian PNI, we designed a structured search strategy grounded on the vaccines listed in the official National Vaccination Schedule\footnote{https://www.gov.br/saude/pt-br/composicao/svsa/pni}. Each vaccine was represented by its official name and expanded with common lexical variations, acronyms, and associated disease terms used in public discourse. For example, the vaccine \textit{BCG} was searched using the central term ``BCG'' and its related disease ``tuberculose'', while the \textit{Polio} vaccine was captured through variations such as ``vip'', ``vop'', ``polio'', ``poliomielite'', and ``paralisia infantil''. This approach was systematically applied across all vaccines, including compound vaccines (e.g., ``penta valente'' for DTP+Hib+HepB), adult boosters (``difteria e tétano''), and respiratory vaccines (e.g., ``influenza'' (``gripe'', ``h1n1'')).

To contextualize the vaccine-related queries and capture diverse framings of public debate, each keyword (denoted as \texttt{\{kw\}}) was combined with a set of intent-specific modifiers reflecting common concerns and narratives surrounding vaccination. These included policy-related terms (e.g., ``Vacina \texttt{\{kw\}} obrigatória''), health risk perceptions (``Perigos da vacina \texttt{\{kw\}}''), adverse outcomes (``Reações adversas vacina \texttt{\{kw\}}'', ``Efeitos colaterais vacina \texttt{\{kw\}}''), mortality claims (``Morte causada por vacina \texttt{\{kw\}}''), long-term consequences (``Sequelas vacina \texttt{\{kw\}}''), and public health monitoring (``Cobertura vacinal \texttt{\{kw\}}''). This query expansion strategy allowed us to capture not only informational content but also controversy-driven and oppositional narratives.

To ensure data consistency and reduce noise, we applied a post-retrieval filter retaining only videos whose titles contained at least one predefined keyword or its variation. Entries without explicit matches were discarded to improve semantic relevance.

\section{Annotation Guidelines}
\label{appendix:guidelines}

The stance categories are defined according to the following criteria:

\begin{itemize}[nosep, leftmargin=*]
\small
    \item \textbf{Favorable:} Explicitly support vaccination, express positive attitudes, share pro-vaccine informational content, highlight benefits, or report positive personal experiences with vaccines.
    \item \textbf{Against:} Explicitly criticize vaccination, present arguments against vaccines, express concerns about adverse effects, promote conspiracy theories, deny scientific evidence, or articulate generalized skepticism toward vaccination science.
    \item \textbf{Inconclusive:} Do not clearly belong to either of the previous categories, deviate from the topic of vaccination, contain ambiguous language, lack sufficient information to infer stance, or are irrelevant to the vaccination debate.
\end{itemize}

\begin{acks}
\small
This work was supported by the Conselho Nacional de Desenvolvimento Científico e Tecnológico (CNPq), the Fundação de Amparo à Pesquisa do Estado de Minas Gerais (FAPEMIG), and the Instituto Nacional de Ciência e Tecnologia em Inteligência Artificial Responsável para Linguística Computacional, Tratamento e Disseminação de Informação (INCT-TILD-IAR). We also acknowledge institutional support from the Universidade Federal de Ouro Preto (UFOP), the Universidade Federal de Minas Gerais (UFMG), and Worcester Polytechnic Institute (WPI).
\end{acks}

\bibliographystyle{ACM-Reference-Format}
\bibliography{mybibfile}

\end{document}